# Change in Tetracene Polymorphism Facilitates Triplet Transfer in Singlet Fission-Sensitized Silicon Solar Cells

Benjamin Daiber[1,§], Sourav Maiti[2,§], Silvia Ferro[1], Joris Bodin[1], Alyssa F.J. van den Boom[3], Stefan L. Luxembourg[4], Sachin Kinge[5], Sidharam Pujari[3], Han Zuilhof[3,6], Laurens D.A. Siebbeles [2*], Bruno Ehrler[1*]

[1] *Center for Nanophotonics, AMOLF, Science Park 104, The Netherlands*

[2] *Optoelectronic Materials Section, Department of Chemical Engineering, Delft University of Technology, Van der Maasweg 9, 2629 HZ Delft , The Netherlands*

[3] *Laboratory of Organic Chemistry, Wageningen University, Stippeneng 4, 6708 WE Wageningen, The Netherlands*

[4] *TNO Energy Transition – Solar Energy, Westerduinweg 3, 1755 LE Petten, The Netherlands*

[5] *Toyota Motor Europe, Materials Research & Development, Hoge Wei 33, B-1913, Zaventem, Belgium*

[6] *School of Pharmaceutical Science and Technology, Tianjin University, 92 Weijin Road, Tianjin, China*

§ Author contributions

B.D. and S.M. contributed equally to this work.

*Corresponding authors

Laurens D.A. Siebbeles: l.d.a.siebbeles@tudelft.nl

Bruno Ehrler: b.ehrler@amolf.nl

## Abstract

Singlet fission in tetracene generates two triplet excitons per absorbed photon. If these triplet excitons can be effectively transferred into silicon (Si) then additional photocurrent can be generated from photons above the bandgap of Si. This could alleviate the thermalization loss and increase the efficiency of conventional Si solar cells. Here we show that a change in the polymorphism of tetracene deposited on Si due to air exposure, facilitates triplet transfer from tetracene into Si. Magnetic field-dependent photocurrent measurements confirm that triplet excitons contribute to the photocurrent. The decay of tetracene delayed photoluminescence was used to determine a triplet transfer time of 215±30 ns and a maximum yield of triplet transfer into Si of ~50 %. Our study suggests that control over the morphology of tetracene during deposition will be of great importance to boost the triplet transfer yield further.

**TOC:**

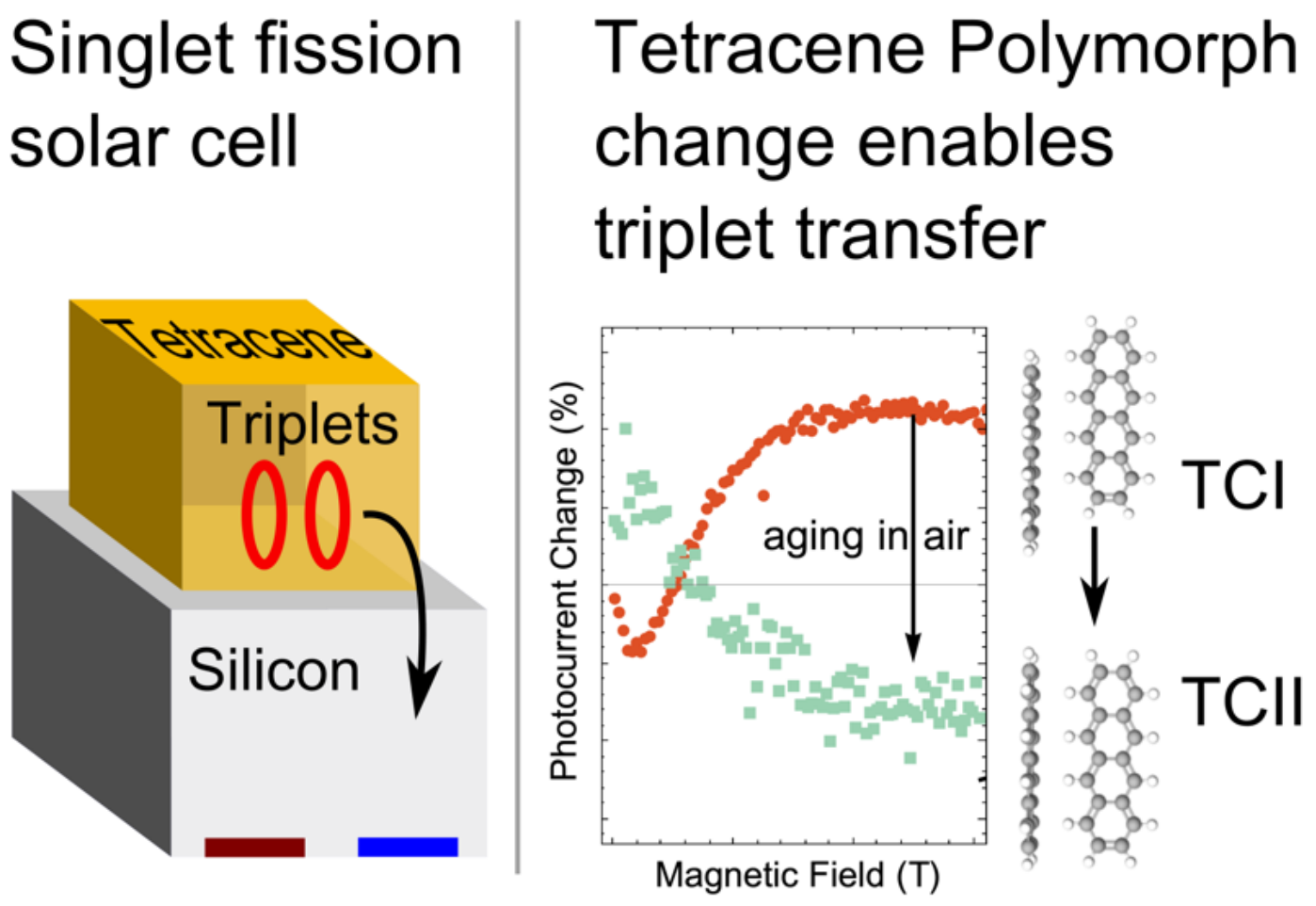

Silicon (Si) is currently the dominating semiconductor material for solar cells, but suffers from several loss mechanisms that reduce its efficiency.[1-2] The largest loss mechanism results from the inefficient utilization of high-energy photons. The additional energy between the Si band gap and the high-energy photons is lost to heat. Sensitizing Si solar cells with a top layer of singlet fission material can reduce this loss, and theoretically even overcome the Shockley-Queisser efficiency limit of ~31% for a single-junction solar cell.[3-14]

Singlet fission is a spin-allowed process of creating two triplet excitons from one singlet exciton that can occur in certain organic semiconductor materials with delocalized $\pi$-orbitals.[5, 15-17] In this paper, we will focus on tetracene, which consists of four benzene rings that are annularly and linearly fused (Figure 1a). Upon absorption of a high energy photon (>2.4 eV) a singlet exciton (bound electron-hole pair) is formed. This singlet exciton ($S_1$) can subsequently be split into two triplet excitons (T) with each roughly half the energy of the singlet exciton. This singlet fission process is mediated by a pair of spin-correlated triplets (TT), based on the kinetic model proposed by Johnson and Merrifield in 1970[18-20]

$$S_0 + h\nu \rightarrow S_1 \rightarrow (TT) \rightarrow T + T \ , \qquad (1)$$

where $S_0$ is the singlet ground state, $h\nu$ the incoming photon energy, and T+T a pair of free triplets. In this model, the rate of singlet fission is determined by the coupling between the $S_1$ and TT states.[21-22] Singlet fission competes with other processes (e.g. radiative and non-radiative recombination and excimer formation), such that some singlet excitons are lost and cannot undergo singlet fission. In tetracene, one absorbed photon leads to close to 2 triplet excitons, as singlet fission is very fast compared to other competing decay channels.[23-24]

In a solar cell architecture where the triplet exciton is transferred into Si, the bandgap of the Si cell has to be smaller than the energy of the triplet exciton state of the singlet fission material. In tetracene, the triplet energy is ~1.25 eV which exceeds the Si bandgap of 1.1 eV, allowing triplet exciton transfer.[24-27] The $V_{oc}$ is determined by the low-bandgap semiconductor Si and the photocurrent from the high-energy photons can be doubled due to singlet fission that eventually generates two electron-hole pairs from the high-energy photons of energy >2.4 eV.

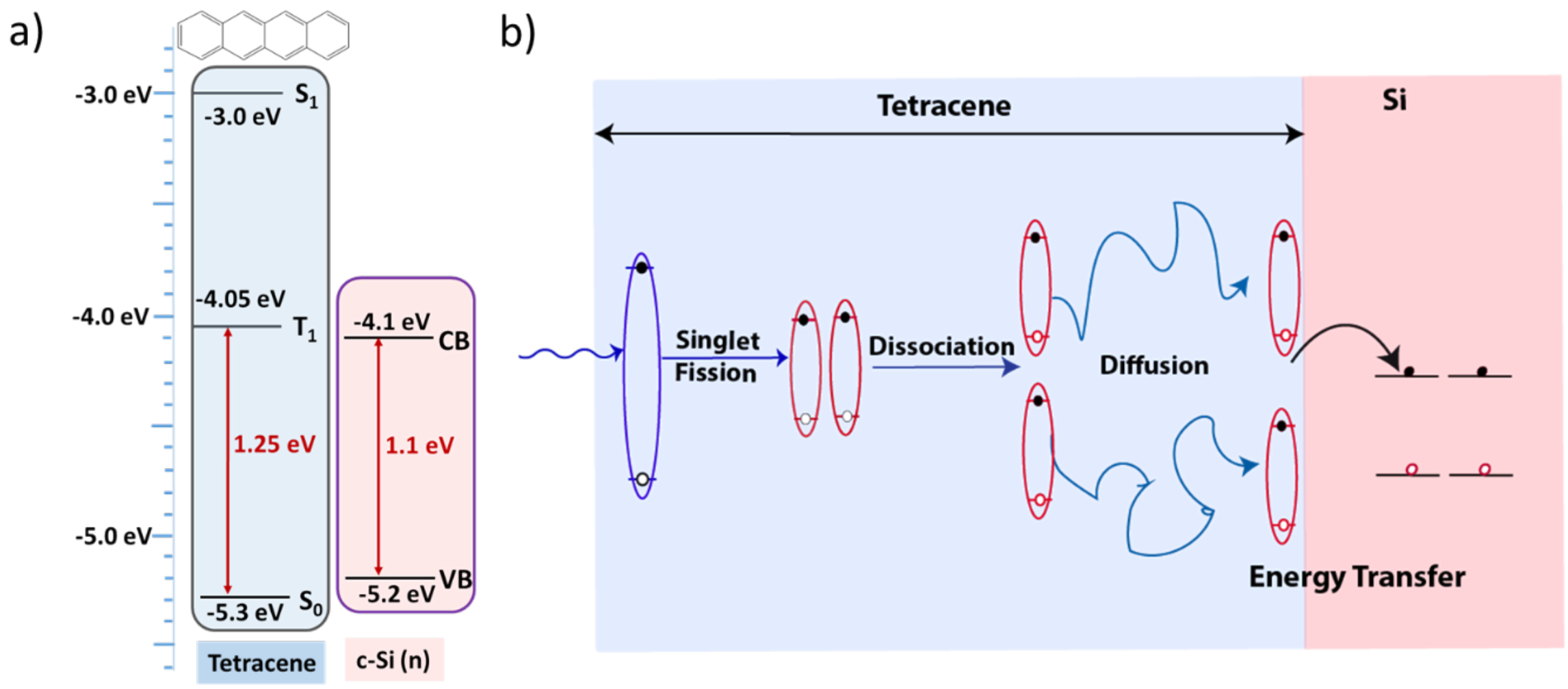

**Figure 1. a)** Energy alignment of tetracene in the ground state ($S_0$), triplet state ($T_1$), singlet state ($S_1$) and Si valence band (VB) and conduction band (CB) from the literature.[24-26] The structure of tetracene is shown at the top. **b)** Schematic of a singlet fission-sensitized Si solar cell. Photons at or above the singlet energy of tetracene are absorbed and create one singlet exciton, which splits into two triplet excitons forming a correlated triplet pair (TT) *via* singlet fission. The TT dissociates into free triplets which can then independently diffuse to the tetracene-Si interface and transfer into Si to generate free charge carriers.

Triplet transfer to Si can happen through energy transfer or charge transfer. In case of energy transfer both electrons and holes arrive in Si concurrently. However, if either electrons or holes are transferred into Si via charge transfer, the remaining countercharges in tetracene have to be extracted by an additional contact.[26] Hence, if triplet energy transfer into Si could be realized the resulting tetracene-Si solar cell would not need an additional charge extracting electrode on top of the tetracene layer. Therefore, energy transfer could in principle enable a simpler solar cell architecture and less added cost to silicon solar cell manufacturing. As the tetracene triplet energy is higher than the Si band gap, transfer into Si is energetically allowed. Figure 1a) shows the ionization energy of the tetracene exciton states and the position of the Si bands. The absolute energy level of the triplet exciton ionization energy with respect to the vacuum level is reported to be in the range of -4.0 eV to -4.3 eV.[24, 26] Figure 1b) shows a schematic of the processes involved in the operation of a singlet fission-sensitized Si solar cell, singlet generation, singlet fission, triplet diffusion, and triplet transfer.

To date, the transfer of triplet excitons from the singlet fission layer to the underlying low-bandgap semiconductor has proven to be the bottleneck for real-world applications. The extraction of triplets directly from tetracene into Si has been investigated by several research groups. Piland *et al.* did not find any evidence of triplet transfer from tetracene into Si upon direct deposition and with a LiF spacer.[28] MacQueen *et al.* reported a small contribution of triplets to the photocurrent upon direct deposition of tetracene on a Si solar cell.[26] The reasons for inefficient triplet transfer could be related to insufficient passivation of the Si surface and the weak coupling between the triplet exciton molecular orbitals and the electronic states in Si. Recently, Einzinger *et al.* unambiguously reported successful triplet transfer from tetracene into Si with 75% efficiency after passivating the Si with a thin (8 Å) dielectric layer of hafnium oxynitride ($HfO_xN_y$) grown through atomic layer deposition (ALD).[29] The ALD-grown interlayer passivates the Si surface and is thin enough to allow the transfer of triplets from tetracene into Si. However, this system is very sensitive to the exact interlayer thickness and composition, and the effect of the tetracene structure remains unclear. The transfer mechanism is still under debate and additional self-passivation effects complicate the interpretation.

Here we report evidence for the triplet exciton transfer in a simpler system, from tetracene into bare Si, after exposure of the tetracene layer to ambient air. We find signatures of triplet exciton transfer in magnetic field-dependent photocurrent measurements and a faster decay of the delayed photoluminescence (PL) from tetracene, indicating triplet exciton quenching. We correlate these changes to a change in tetracene morphology as seen in X-ray diffraction (XRD) spectra that show the conversion of polycrystalline tetracene from polymorph I (TCI) to polymorph II (TCII), which exhibits a faster singlet fission rate.[30-31] We propose that the change of tetracene polymorph is important for the observed triplet transfer into Si solar cells.

Measuring the effect of triplet excitons on the photocurrent of a solar cell is the most direct way of measuring the transfer of triplet excitons, and is most relevant for the real-world application of a singlet fission-sensitized Si solar cell. The final goal is to increase the Si photocurrent from transferred triplet excitons. However, both singlet and triplet excitons can contribute to the photocurrent. Therefore, it is important to prove whether photocurrent originates from triplet versus singlet excitons. To distinguish between singlet and triplet exciton transfer we exploit the behavior of singlet fission under a magnetic field (see Figure 2). Under a magnetic field of 300 mT singlet fission in tetracene becomes less efficient, resulting in a lower triplet exciton population compared to the situation without the magnetic field.[20] If the photocurrent from Si has the same magnetic field dependence as the triplet population (blue curve in Figure 2a) we conclude that there is transfer of triplet excitons into Si. The photocurrent is caused prevalently by triplet transfer; the opposite magnetic field dependence (yellow in Figure 2a) would indicate that the photocurrent is dominated by singlet transfer or radiative transfer. The relationship between magnetic field and singlet fission efficiency (or singlet/triplet populations) is not monotonic, below 50 mT there is a small dip in the opposite direction as described by Merrifield *et al.* (see also Figure 2a).[20] This characteristic curve also allows us to exclude any other effects that the magnetic field could have on the photocurrent, like a displacement of the sample, induced currents at the contacts or sample degradation over time.

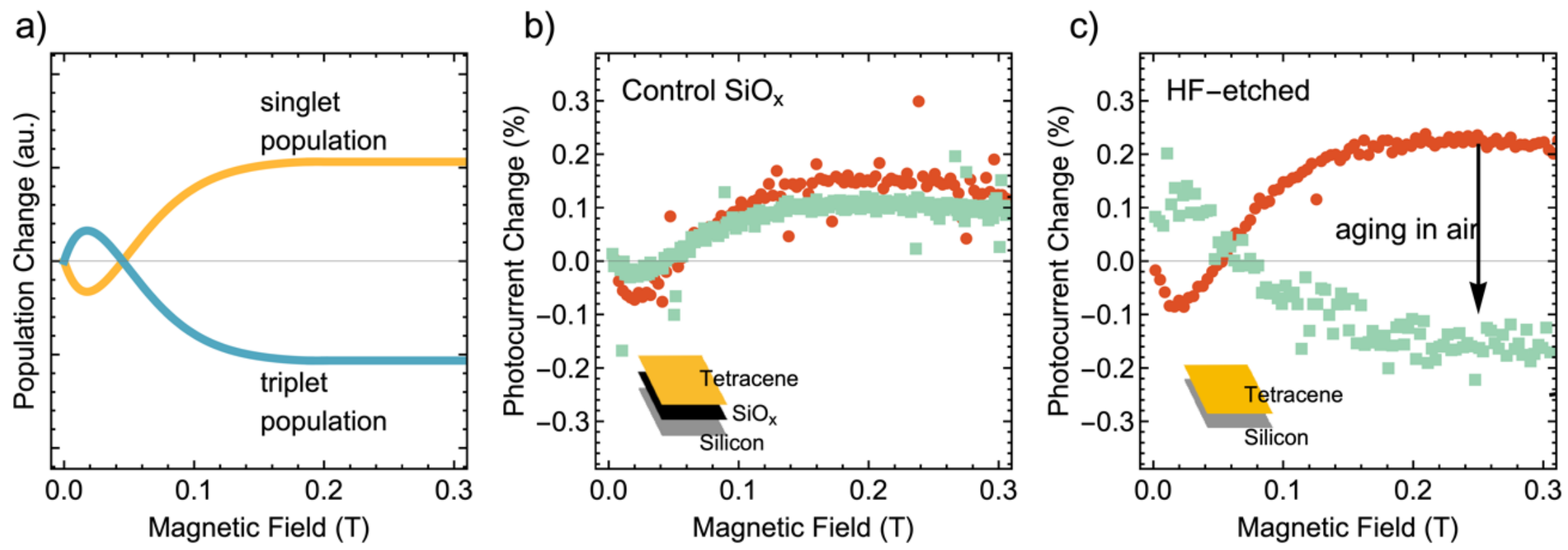

**Figure 2. a)** Schematic for the behavior of singlet and triplet population in tetracene for photocurrent as a function of the magnetic field. **b/c**) Magnetic field-dependent photocurrent measurements for **b)** Si/$SiO_x$/tetracene (control), and **c)** HF-Si/tetracene, both before (red curve) and after (green curve) exposure to air. For both samples **b)** and **c)** the positive change in photocurrent can be attributed to the dominant contribution of singlets. Aging the HF-etched sample in air flips the curve and leads to a triplet curve, indicating that triplets are transferred and contributing to the Si photocurrent.

We fabricated Si solar cells with an additional tetracene singlet fission top layer. The solar cells are then encapsulated in an inert $N_2$ atmosphere between two glass slides to keep oxygen and moisture out. Between the tetracene layer and the Si solar cell we used different interlayers for reference measurements and to gain insight into the transfer mechanism. We then measured the photocurrent as a function of an externally applied magnetic field as described above. Figure 2b) shows the magnetic field dependent photocurrent of solar cells with an insulating interlayer of ~2 nm $SiO_x$, which shows no signature of triplet exciton transfer. A thick (~80 nm) $Si_3N_4$ (SiN) interlayer shows the same blocking behavior, as shown in Figure S1a) in the Supporting Information. The photocurrent follows the curve we would expect for singlet excitons, indicating that the singlet excitons contribute to the photocurrent. Utilizing a HF-etch to remove the blocking layer and enabling direct contact between tetracene and Si (HF-Si/tetracene) does not change this behavior, as seen in Figure 2c) (red curve), which is in line with earlier reports.[28] The photocurrent still follows the singlet exciton population, and no evidence for triplet transfer is observed.

Si/$SiO_x$/tetracene and HF-Si/tetracene samples were then stored in air under ambient conditions in the lab for five days and re-measured (Figure 2b) and c)). The magnetic-field dependence of the photocurrent curve for the HF-Si/tetracene solar cell, shown in Figure 2 c), reverses for the air-exposed sample, closely following the characteristic shape for a triplet exciton population, which is strong evidence for triplet exciton transfer. If we encapsulate the solar cell and store it in air, we also observe the triplet curve, although its emergence is then much slower, i.e. after six weeks, as shown in Figure S1b) (see Supporting Information), indicating that eventually air enters the encapsulation. If the HF-Si/tetracene solar cell is stored under a dry nitrogen atmosphere in the glovebox (<10 ppm $O_2$, <1 ppm $H_2O$), we instead observed the singlet curve, which was retained after six weeks (Figure S1c), Supporting Information). The strong difference in magnetic-field photocurrent behavior between the air-exposed and nitrogen-stored samples indicates that air-exposure plays a crucial role in enabling successful triplet transfer to Si.

In Figure 2c) the decrease in photocurrent at high field is around 0.2%, which is comparable to silicon-tetracene solar cells with $HfO_xN_y$ interlayers.[29] In that study the self-passivation in Si, due to improved surface screening by charge carriers at the Si interface, causing an increased photocurrent.[29] This self-passivation can lead to an overestimation of the contribution of triplet exciton injection, and the effects of triplet excitons and self-passivation were separated by a strong background illumination. We performed similar experiments to investigate self-passivation of Si in our samples by using a strong (100 W) xenon light source with red light below the absorption onset of tetracene but above the absorption onset of Si. This allows us to inject charge carriers directly in Si that cannot have originated in tetracene. We did not see an influence of this additional light on the photocurrent change under the magnetic field after correcting for the additional bias current, so we can exclude large influences from self-passivation in the Si solar cell (See Figure S2, Supporting Information). Therefore, we can conclude that exposure to air leads to triplet transfer from tetracene into Si. The following experiments will offer additional evidence for triplet exciton transfer and insight into the dynamics and mechanism.

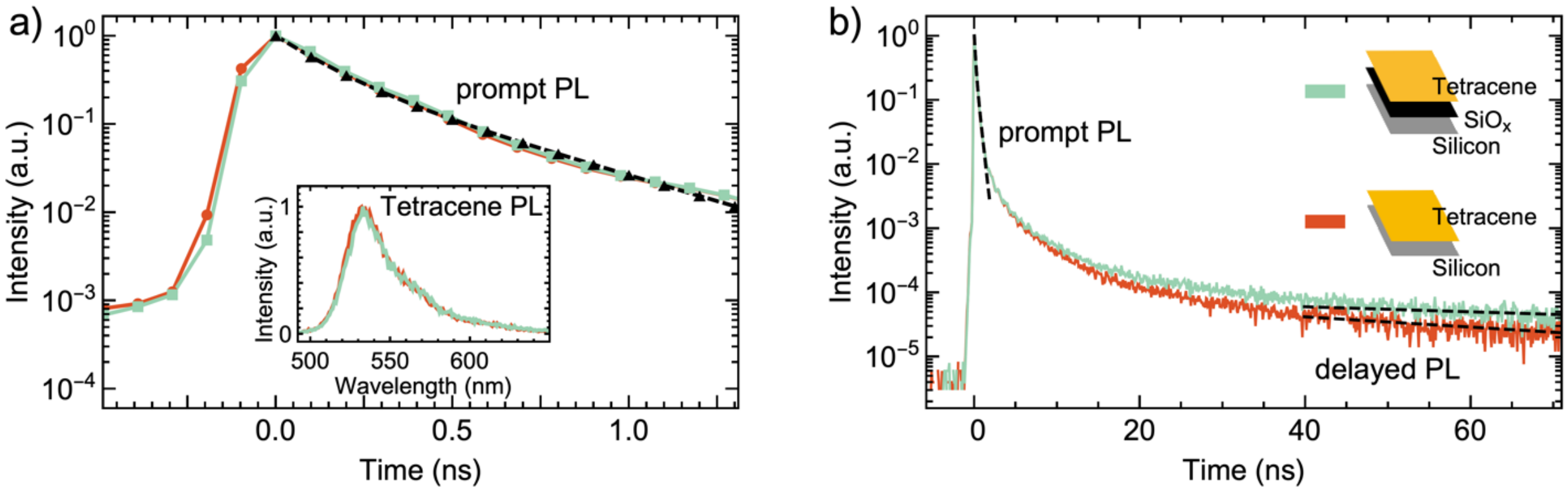

**Figure 3.** The PL decay traces in $Si/SiO_x$/tetracene and HF-Si/tetracene solar cells after air exposure. The dotted black line represents the output from the kinetic model; **a)** short-time (prompt) PL shows no difference in singlet fission time and efficiency between samples; **b)** long

time (delayed) PL shows faster decay for the HF-Si/tetracene solar cell, which we attribute to triplet transfer into Si. The inset in **a)** shows the PL spectra of both samples.

To investigate the mechanism, timescale, and yield of the transfer process of triplets into Si, we measured the tetracene PL decay both in the solar cells and tetracene deposited on Si wafers. The inset in Figure 3a), compares the PL spectra of Si/$SiO_x$/tetracene and HF-Si/tetracene solar cells after exposure to air, showing the characteristic tetracene 0-0 emission peak at 535 nm with a shoulder at 580 nm due to the 0-1 transition, and a broad defect emission around 615 nm. This defect emission arises from structural defects in tetracene during the vacuum evaporation process.[32-33] The PL spectra are the same for both the samples upon air exposure, thus the PL spectra show no evidence of additional trap states from the aging process in samples with and without the $SiO_x$.

The PL in tetracene originates from singlet exciton emission. The PL decay shows a fast initial component due to singlet fission at short times (<1 ns, prompt PL) and a long-lived delayed PL arising from the triplet-triplet annihilation to singlet excitons at later times (>40 ns, delayed PL). The prompt PL decay is identical for the aged samples with and without the $SiO_x$ blocking layer, showing that the singlet fission rate is not affected by the blocking layer (Figure 3a). However, the delayed PL component was faster in the HF-Si/tetracene solar cell upon air exposure, which provides additional evidence for the depopulation of triplets in tetracene caused by triplet transfer into Si (Figure 3b). Since triplet excitons disappear from tetracene because they are transferred into Si, triplet-triplet annihilation is reduced which in turn reduces the delayed PL intensity. The PL lifetime measurements have been reproduced with solar cell samples having a SiN blocking layer (Figure S3, Supporting Information) showing no evidence for triplet transfer, just like in the $SiO_x$ samples. The samples with the SiN blocking layer and HF-Si have been measured at different spots of the solar cell (Figure S4, Supporting Information), and showed no dependence on measurement position on the samples. To reconfirm the results PL decay measurements were performed in tetracene deposited on Si wafers (i.e., not a full solar cell) with and without the $SiO_x$ blocking layer, and similar results were obtained (See Figure S5, Supporting Information). This ensures that the observed PL dynamics are a characteristic of the HF-Si/tetracene interface, and that it is not influenced by the presence of other solar cell components. Together with the magnetic field-dependent photocurrent measurements, we therefore correlate the faster decay in the delayed PL to triplet quenching due to the triplet transfer process from tetracene into Si.

What is the mechanism activating triplet transfer in the HF-Si/tetracene samples after aging in air? The activation could originate from either a change in the tetracene or of the HF-Si interface or of both. We deployed X-Ray diffraction (XRD) to detect changes in the tetracene morphology and X-ray Photoemission Spectroscopy (XPS) to investigate changes on the HF-Si surface. Two different polymorphs, created by heated (TCI) or cooled (TCII) substrates can form during tetracene deposition.

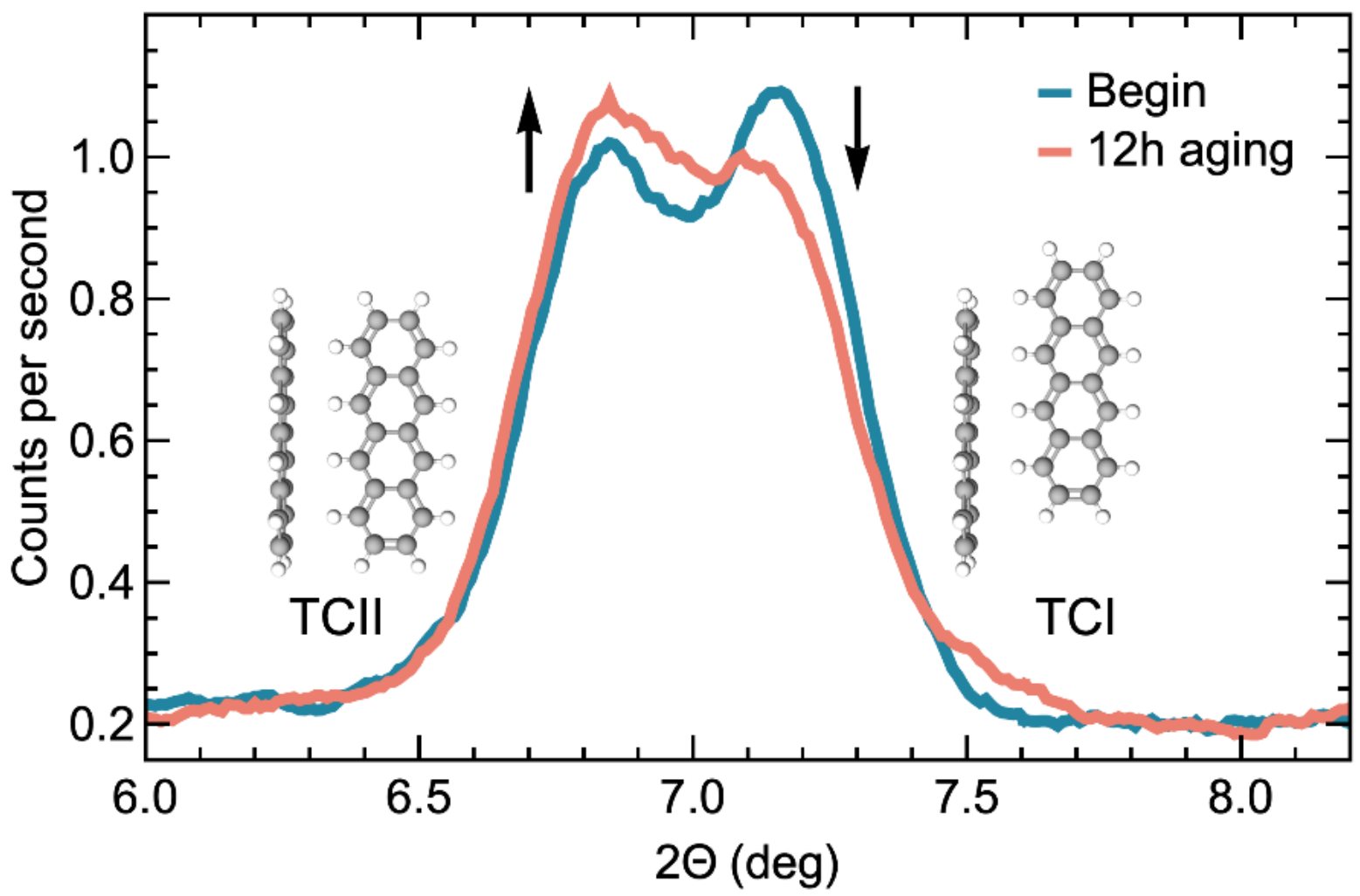

**Figure 4.** XRD of tetracene deposited on fresh HF-Si/tetracene and after 12 h of air exposure. We observe a conversion of TCI to TCII.

The TCII polymorph has different packing with increased distance along the c-axis compared to TCI, resulting in a lower diffraction angle in XRD along the (00c) diffraction.[31] Before air exposure, the XRD spectra shows the presence of both polymorphs with slightly more TCI ($2\Theta = 7.3°$) compared to TCII ($2\Theta = 6.9°$) as seen in Figure 4. However, after air exposure, the ratio reversed with more TCII compared to TCI, suggesting a change in polymorphism in tetracene. The two polymorphs have different intermolecular coupling strengths due to a difference in molecular orientations that leads to a faster singlet fission rate in TCII compared to TCI as reported by Arias *et al.*[30] We also observe a change in singlet fission rate, as the prompt PL decay of tetracene deposited on quartz becomes faster after air exposure (Figure S6, Supporting Information).

To confirm that the change responsible for triplet transfer is the aging of tetracene and not the aging of the HF-Si surface we exposed HF-Si samples to air for different amounts of time, to grow an $SiO_x$ layer with various thicknesses. We measured XPS to confirm the growth of this $SiO_x$ overlay by monitoring the Si-$O_x$ peak in the Si 2p photoelectron emission narrow scan (Figure S7a) and b), Supporting Information). On these samples we deposited fresh tetracene and measured the PL-decay. Triplet transfer was not observed in these samples (Figure S7c), Supporting Information) confirming that triplet transfer is not associated with the growth of a $SiO_x$ layer, and the ageing and subsequent change in polymorphism of tetracene is related to the triplet transfer.

Triplet transfer via a direct Dexter-type mechanism is dependent on the overlap between the triplet exciton wavefunction of tetracene and of the electron and hole wavefunctions at the Si surface.[34] This coupling will change depending on the distance and orientation of the tetracene molecules with respect to the Si surface. Therefore, the change in the orbital coupling in going from TCI to TCII and its effect on the triplet transfer efficiency is likely crucial and needs to be investigated further theoretically. A recent report by Niederhausen *et al.* that deployed near-edge X-ray absorption fine structure (NEXAFS), XPS and density functional theory (DFT) calculations also suggests that different orientations at the interface exist and could lead to a change in transfer efficiency.[35]

To extract the triplet transfer rate and transfer efficiency we model the PL decay data considering the singlet fission process as depicted in Figure 5a).[28, 30, 36] The singlet fission process in tetracene (with rate: $k_{SF}$) competes with the radiative decay ($k_{Rad}$) through the formation of triplet pairs, which can then dissociate ($k_{Diss}$) to form free triplets or fuse back ($k_{TT}$) to create an excited singlet state. The free triplets can decay with the triplet lifetime ($k_{Trip}$), or regenerate the triplet pair state through triplet-triplet annihilation ($k_{TTA}$). TTA results in the delayed PL from tetracene and determines the triplet lifetime.[28, 37-38] The populations of the $S_1$, TT and T states can be determined by solving the coupled differential equations as detailed in the Supporting Information. The $S_1$ population is plotted against the measured PL decay traces in Figure 3 (dotted lines) for the prompt and delayed PL. The rate constants described above were determined by solving the differential equation for HF-Si/tetracene before and after exposure to air and are compared to the data. The rate constants obtained are in agreement with the literature as shown in Table S1, Supporting Information. The singlet fission time was determined to be 175 ± 10 ps, which is in good agreement with reported values of ~75 - 200 ps depending on the crystallinity, grain size and preparation conditions.[28, 30, 33, 36, 38-39] The faster decay of the delayed PL can be reproduced with the kinetic model by incorporating an additional triplet transfer process to Si without changing the other rate constants. The model based on these kinetic equations reproduces the data for prompt (until 1 ns) and delayed PL (>40 ns) well, but fails to describe the decay at intermediate times between prompt and delayed PL. This is most likely due to additional effects of triplet pair diffusion, which have been explained by modelling that leads to a $t^{-3/2}$ dependence of the PL decay.[40] Our intermediate PL decay data is also described by this function (Figure S8, Supporting Information).[40]

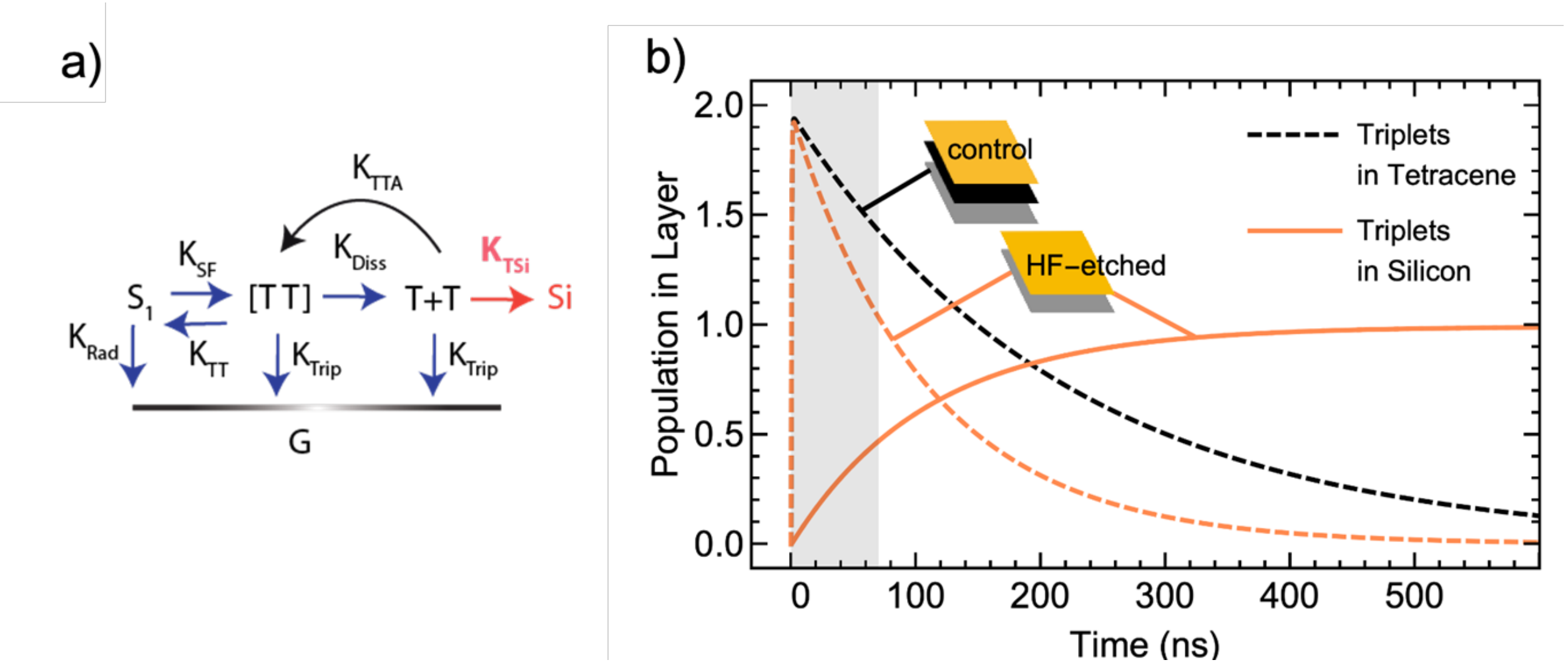

**Figure 5. a)** Schematic representation of the kinetic model used to determine the triplet transfer efficiency; **b)** the triplet population in Si/$SiO_x$/tetracene (control), HF-Si/tetracene and in HF-Si after air exposure predicted from our model as a function of time. The grey area shows our experimental range of the PL decay measurements (70 ns).

From the kinetic model, the triplet lifetime is estimated to be ~220 ± 30 ns in both samples and the triplet transfer time into Si is ~ 215 ± 30 ns in the aged sample. Figure 5b) shows how the triplet population varies with time with a faster decay of the triplet population in HF-Si due to triplet transfer into Si. The growth of the triplet population in HF-Si corresponds to ~25% of triplet transfer within our experimental measurement window of 70 ns. Extrapolation to longer times shows ~50% of triplet transfer at ~500 ns. Our findings show that efficient triplet transfer can be obtained when the percentage of TCII is increased. Thus, to achieve even higher triplet transfer yield into HF-Si, control over the morphology appears to be crucial, so that TCII becomes the predominant polymorph. This also suggests that apart from the complex $HfO_xN_y$ interlayers used before, we can also exploit the tetracene orientation itself for efficient triplet transfer.[29]

In summary, we have shown that a change in the dominant tetracene polymorph by air exposure facilitates triplet transfer from tetracene into Si. The triplet transfer process is confirmed through magnetic field-dependent photocurrent measurements, and the timescale of triplet transfer is obtained from delayed PL decay measurements. We find that the transition from the tetracene polymorph TCI to TCII is essential for efficient triplet transfer. This suggests that the orientation (w.r.t. the surface), and the packing of the singlet fission molecule is crucial for an efficient triplet transfer process. Future research should focus on an optimal alignment of tetracene molecules by preparing a pure TCII polymorph on Si and potential combination of both interlayers and polymorph control, which could lead to the optimal triplet transfer efficiency. This could then enable cheap manufacturing of singlet fission-sensitized Si solar cells.

**Associated Content: Supporting Information**

Sample preparation and experimental details, additional magnetic field-dependent photocurrent measurements, XPS, additional PL decay measurements.

**Notes**

The authors declare no competing financial interest.

**Acknowledgements**

This research received funding from the Netherlands Organization for Scientific Research (NWO) in the framework of the Materials for sustainability and from the Ministry of Economic Affairs in the framework of the PPP allowance.

# Change in Tetracene Polymorphism Facilitates Triplet Transfer in Singlet Fission-Sensitized Silicon Solar Cells

Benjamin Daiber[1,§], Sourav Maiti[2,§], Silvia Ferro[1], Joris Bodin[1], Alyssa F.J. van den Boom[3], Stefan L. Luxembourg[4], Sachin Kinge[5], Sidharam Pujari[3], Han Zuilhof[3,6], Laurens D.A. Siebbeles [2*], Bruno Ehrler[1*]

[1]*Center for Nanophotonics, AMOLF, Science Park 104, The Netherlands*

[2]*Optoelectronic Materials Section, Department of Chemical Engineering, Delft University of Technology, Van der Maasweg 9, 2629 HZ Delft, , The Netherlands*

[3] *Laboratory of Organic Chemistry, Wageningen University, Stippeneng 4, 6708 WE Wageningen, The Netherlands*

[4] *TNO Energy Transition – Solar Energy, Westerduinweg 3, 1755 LE Petten, The Netherlands*

[5]*Toyota Motor Europe, Materials Research & Development, Hoge Wei 33, B-1913, Zaventem, Belgium*

[6]*School of Pharmaceutical Science and Technology, Tianjin University, 92 Weijin Road, Tianjin, China*

[§] Author contributions

B.D. and S.M. contributed equally to this work.

*Corresponding authors

Laurens D.A. Siebbeles: l.d.a.siebbeles@tudelft.nl

Bruno Ehrler: b.ehrler@amolf.nl

## Experimental section

### Solar cells

The solar cells are Interdigitated Back Contact (IBC) silicon solar cells with a silicon pyramid antireflection layer and a ~80 nm SiN passivation and antireflection layer. We cut the 4 cm × 4 cm solar cells in three stripes with a laser cutter. We use a wire bonder to contact the back-side contacts to contact pads for the photocurrent measurements.

### Silicon Wafers

Silicon Wafers were purchased from Siegert Wafer Gmbh. We used <111> FZ-silicon, n-doped (Ph) with a resistivity of 1-5 Ohm-cm, Double side polished with a thickness of 0.28 mm. We dice the wafers using a laser cutter.

### HF etching of solar cells

We dripped concentrated hydrofluoric acid (HF) solution (40%, Sigma-Aldrich, as received) onto the top surface of the silicon solar cell with a pipette. In this way the HF solution does not contact the metallic back contacts. After 10 min of etching the wettability decreases dramatically, meaning that at this point we have etched away the SiN which has a lower contact angle than the bare Si surface. We then removed any of the remaining HF solution from the surface by dipping the samples sequentially twice in deionized water baths. Immediately afterwards, the sample was transferred into a nitrogen-filled glovebox.

### Tetracene deposition:

We deposited 200 nm of tetracene onto the silicon solar cells, and 35 nm and 200 nm onto the bare silicon samples. The evaporation was done inside an thermal evaporator (Angstrom Engineering

Inc.), at a base pressure below $7{\cdot}10^{-7}$ mbar. Tetracene was purchased from Sigma-aldrich (99.99% purity) and used as is. The deposition was 1 Å/s in all cases. Encapsulation was also done inside the nitrogen-filled glove box, using two glass slides, a rubber gasket and silicone glue. None of the samples were exposed to UV light during storage and measurement.

**Magnetic-field dependent photocurrent measurements:**

We measured the magnetic-field depedent photocurrent using a home-built setup. The magnetic field is applied by an electromagnet, made up by two Helmholtz coils and calibrated using a Hall effect sensor. The magnetic field is applied by sending a current of up to 5 Amperes through the coils, resulting in a magnetic field of up to 0.35 T. The field is oriented parallel to the sample surface. The excitation source is a 520 nm diode laser, installed in a Thorlabs temperature controlled laser housing. The cw laser power is around 10 mW with a laser spot size of approx. 1 mm. The photocurrent is measured with a Keithly 2636A source-measure unit. After each measurement at a certain magnetic field we perform a reference measurement at zero magnetic field. For the self-passivation measurements we added a 100 W xenon lamp with a 550 nm longpass filter, to only excite the silicon sample.

**Photoluminescence studies:**

Photoluminescence (PL) spectra and decay kinetics were monitored in Lifespec-ps (Edinburgh Instruments) upon 404 nm laser pulse (pulse width ~100 ps) excitation with a repetition rate of 200 kHz. The tetracene-deposited substrates were encapsulated inside a custom-made sample holder inside a glovebox prior to the measurement. Afterwards, the samples were exposed to air and encapsulated again inside the glovebox to investigate the effect of air exposure on PL decay.

**X-Ray diffraction:**

XRD was measured on a Bruker D2 Phaser using a Cu Tube with 1.54 Å at 10 mA and 30 kV as a source. We used a Lynxeye detector in 2Theta mode with a scan speed of 420 s per measurement. Each hour we measure the diffraction spectra once. The sample was rotating at 10 /min and kept inside the instrument during the whole measurement. We use a moving average over 21 points to smooth the curve.

**X-Ray Photoelectron spectroscopy:**

XPS measurements were performed with a JEOL JPS-9200 photoelectron spectrometer, using a 12 kV and 20 mA monochromatic Al Kα source. The analyzer pass energy was 10 eV, and the take-off angle between sample and detector was set at 10$^{o}$. Data was analyzed using the CasaXPS program, version 2.3.18PR1.0. To measure the samples, the silicon surfaces with tetracene layer were transferred in an air-tight container from a glovebox ($O_2$ < 0.01 ppm, $H_2O$ < 0.01 ppm) to the XPS, to minimize air exposure during transfer. In the XPS, a first dummy measurement taking ~100 min was performed, to heat up the sample stage. This, in combination with the high vacuum in the measurement chamber ($10^{-5}$ - $10^{-6}$ mbar), led to the sublimation of tetracene from the sample, leaving the bare silicon surface behind. The level of silicon surface oxidation was then assessed by performing narrow scans on the silicon 2p peak. The relative surface area of the Si 2p peak at ~103 eV was taken as a measure for surface oxidation.

## Magnetic-field dependent photocurrent

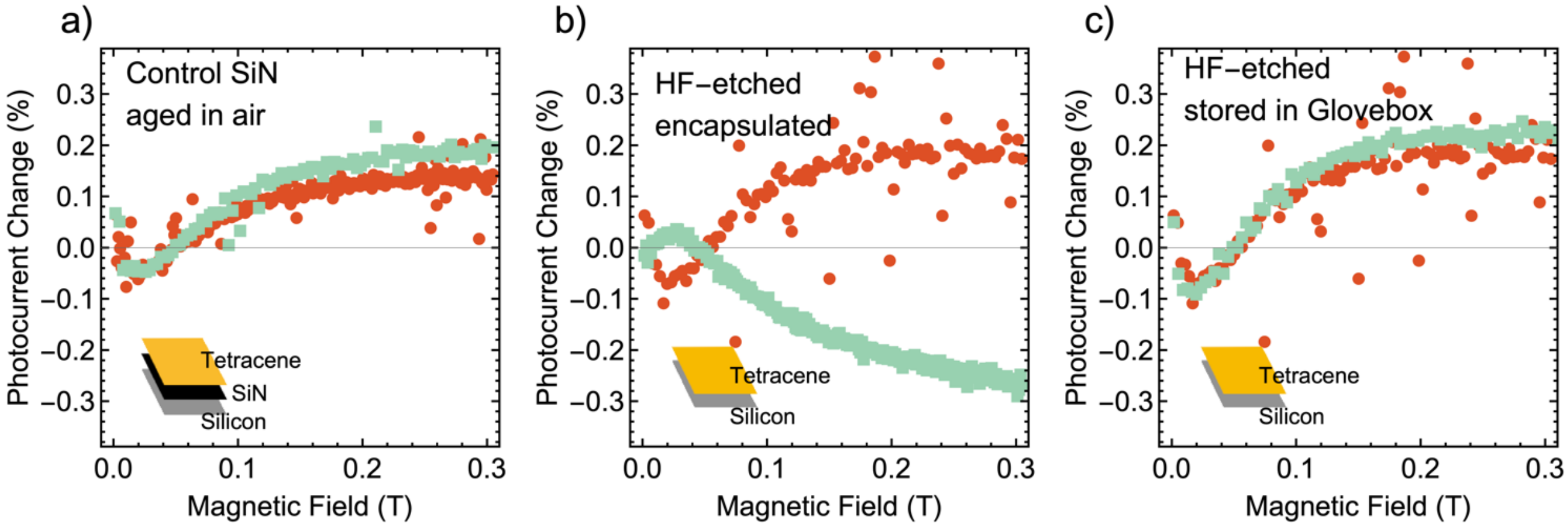

**Figure S1.** Magnetic field-dependent photocurrent in **a)** Si/SiN/tetracene solar cell before (green curve) and after aging (red curve) in air with no change in triplet transfer behavior. **b)** Encapsulated HF-Si/tetracene silicon solar cell stored in air for six weeks. The photocurrent change flips from singlet to triplet curve, and **c)** HF-Si/tetracene solar cell stored under nitrogen atmosphere in the glovebox for six weeks without changes to the triplet transfer.

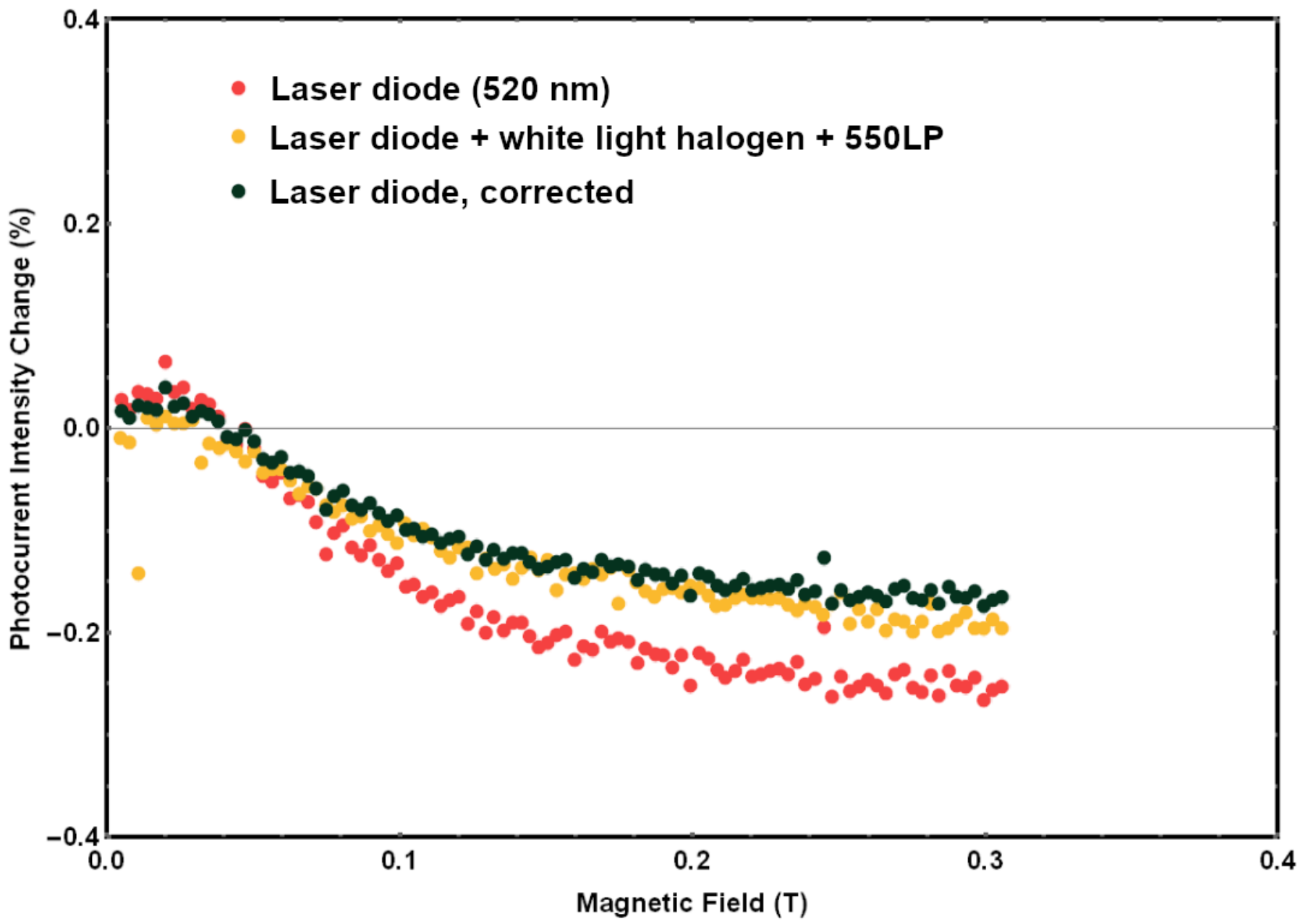

**Figure S2.** Change in photocurrent of HF-Si/tetracene solar cell under magnetic field with different light sources. Illumination with the green laser (red curve), illumination with green laser and red light (yellow), and correcting the red curve with the formula described below. If accounted for the additional charge carriers, both curves are on top of each other and we can conclude that there is a negligible effect of self passivation from the additional charge carriers in silicon.

To investigate whether the self-passivation of additional charge carriers in silicon can lead to an overestimation of the magnetic field effect, we add an additional red light with energy below the tetracene bandgap so that the light is only absorbed in the silicon layer.

The change in photocurrent with magnetic field can be written as $\Delta I(B) = \frac{I(B)-I(B=0)}{I(B)}$. If we add different light sources, each light source will add current that can also be dependent on the magnetic field. In our experiment we added red light (leading to current $I_R(B)$) to the green laser (current $I_G(B)$). The change in photocurrent then becomes

$$\Delta I(B) = \frac{I_R^{Tc}(B) + I_R^{Si}(B) + I_G^{Tc}(B) + I_G^{Si}(B) - I_R^{Tc}(B=0) - I_R^{Si}(0) - I_G^{Tc}(0) - I_G^{Si}(0)}{I_G(0) + I_R(0)}$$

Since tetracene does not absorb in the red, we set $I_R^{Tc}(B) = 0$. The current generated in silicon directly does not depend on the magnetic field, leading to $I^{Si}(B) = I^{Si}(0)$, leaving us with:

$$\Delta I(B) = \frac{I_G^{Tc}(B) - I_G^{Tc}(0)}{I_G(0) + I_R(0)}$$

We measure the current from only the red light at zero field and then correct the magnetic field curve with the formula above.

## Photoluminescence

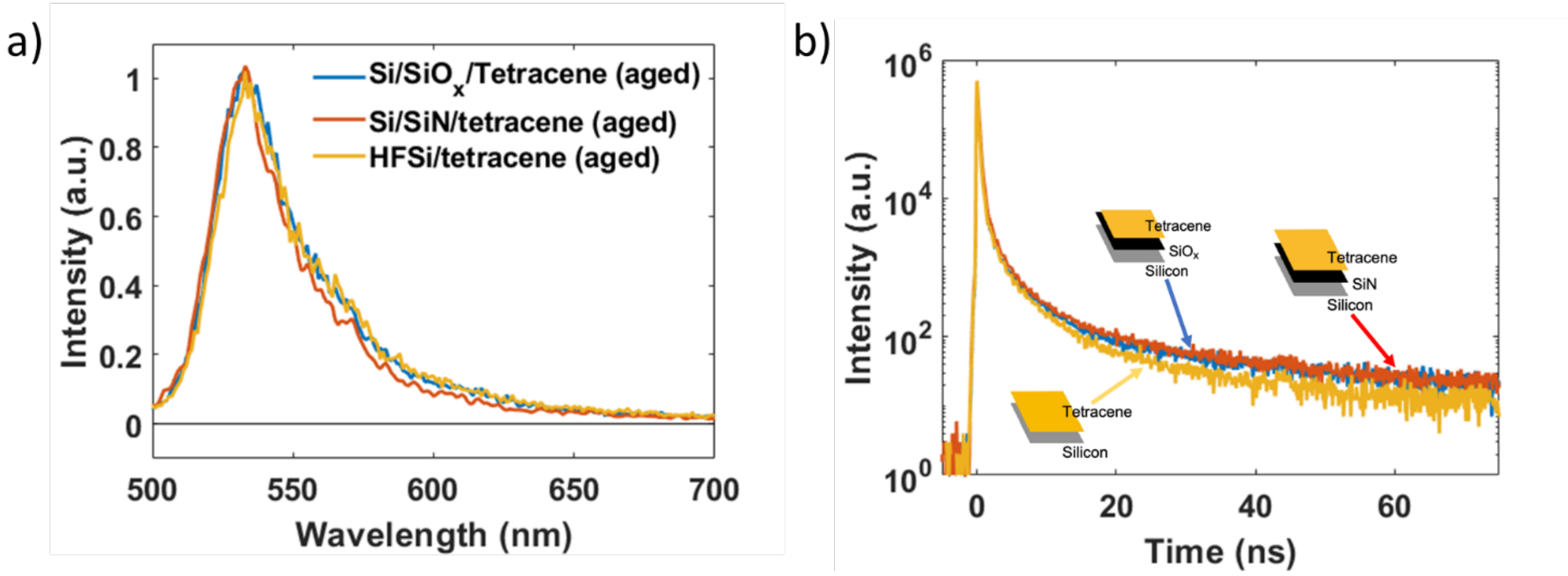

**Figure S3. a)** PL spectra of tetracene show the absence of degradation and additional trap states, and **b)** decay traces for tetracene deposited on Si/SiO$_x$, Si/SiN and HF-Si solar cells, showing no difference between slope of the long-time decay for the SiN and $SiO_x$ blocking layers and triplet quenching in fresh HF-Si.

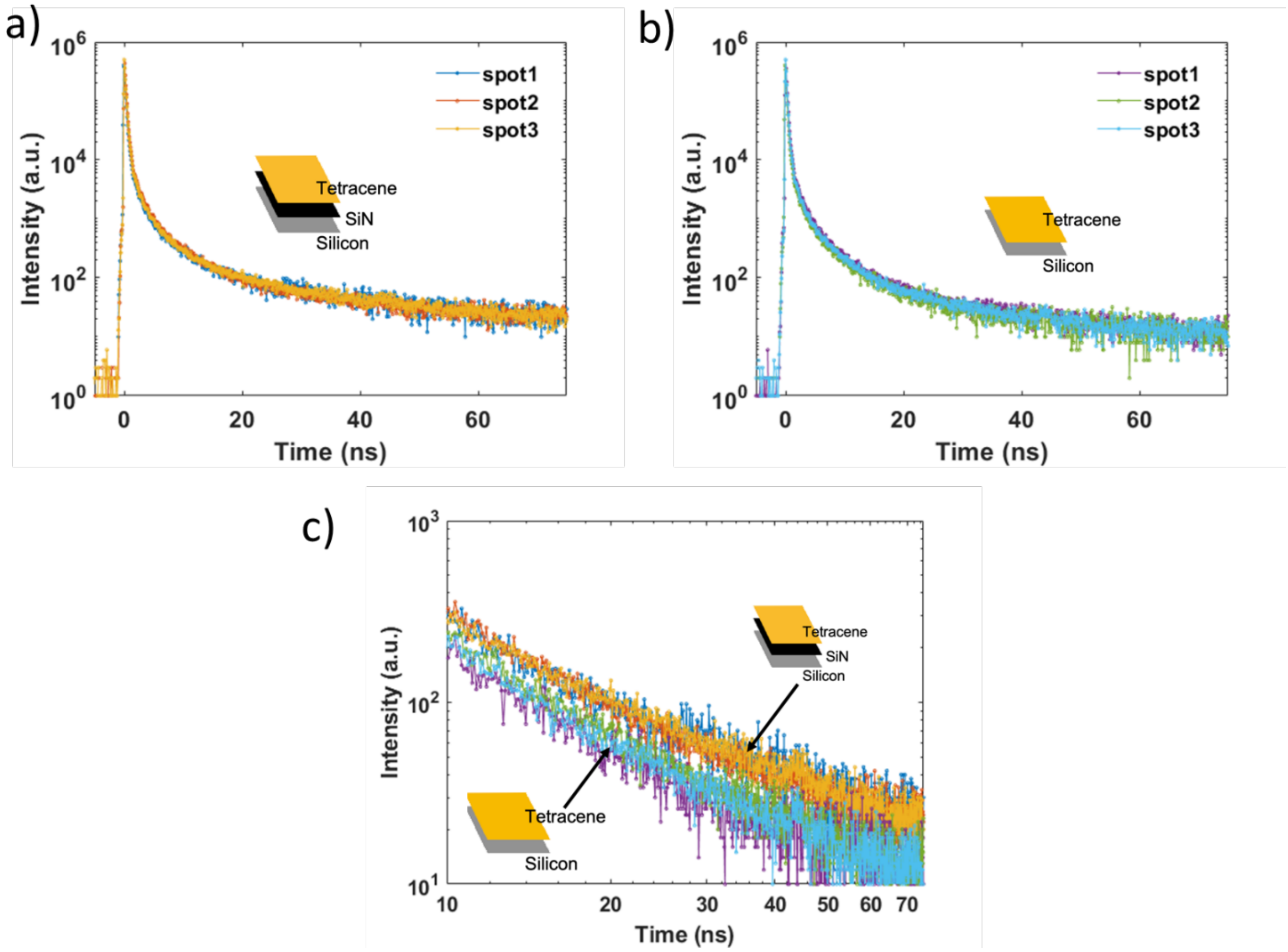

**Figure S4.** PL decay traces for **a)** Si/SiN and **b)** HF-Si solar cells sensitized with tetracene excited on different spots on the solar cell surface. **c)** Comparison of decay traces in **a)** and **b)** together to show the reproducibility of the data in different spots and measurements.

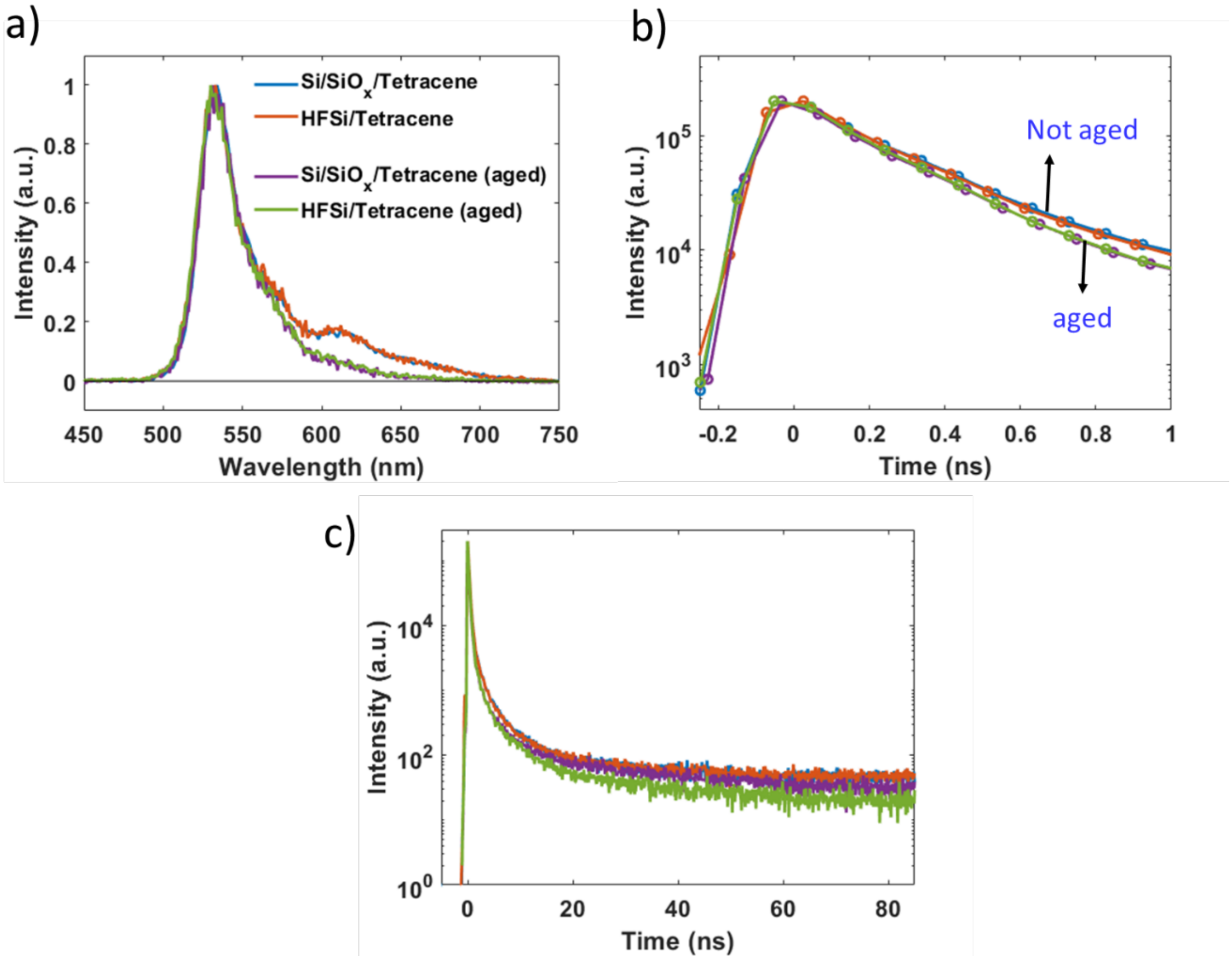

**Figure S5. a)** PL spectra, **b)** prompt and **c)** delayed PL for tetracene deposited on Si/$SiO_x$ and HF-Si wafers (thus not solar cells as in the main text) with and without air exposure. The PL spectrum shows some difference in the defect emission upon expose to air, contrary to the solar cells samples. The decay dynamics show the same effect as the solar cell samples. The initial singlet decay gets equally faster upon air exposure in both Si/$SiO_x$ and HF-Si, implying faster SF rate due to the increased concentration of TCII. The decay of delayed PL is faster in tetracene deposited on HF-Si (green curve) compared to the Si/$SiO_x$ (violet curve) substrate, consistent with triplet exciton transfer.

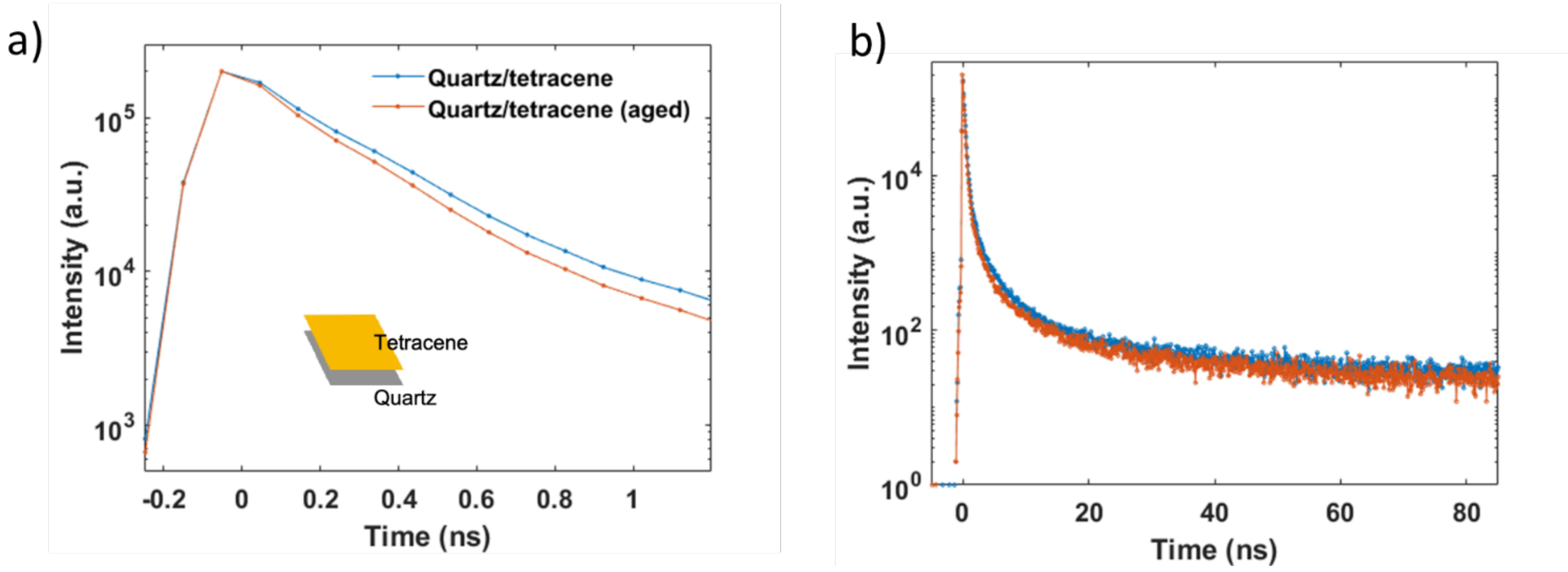

**Figure S6. A)** Prompt and **b)** delayed PL of tetracene deposited on quartz. The initial fast decay after air exposure can be attributed to an interchange of TCI to TCII.

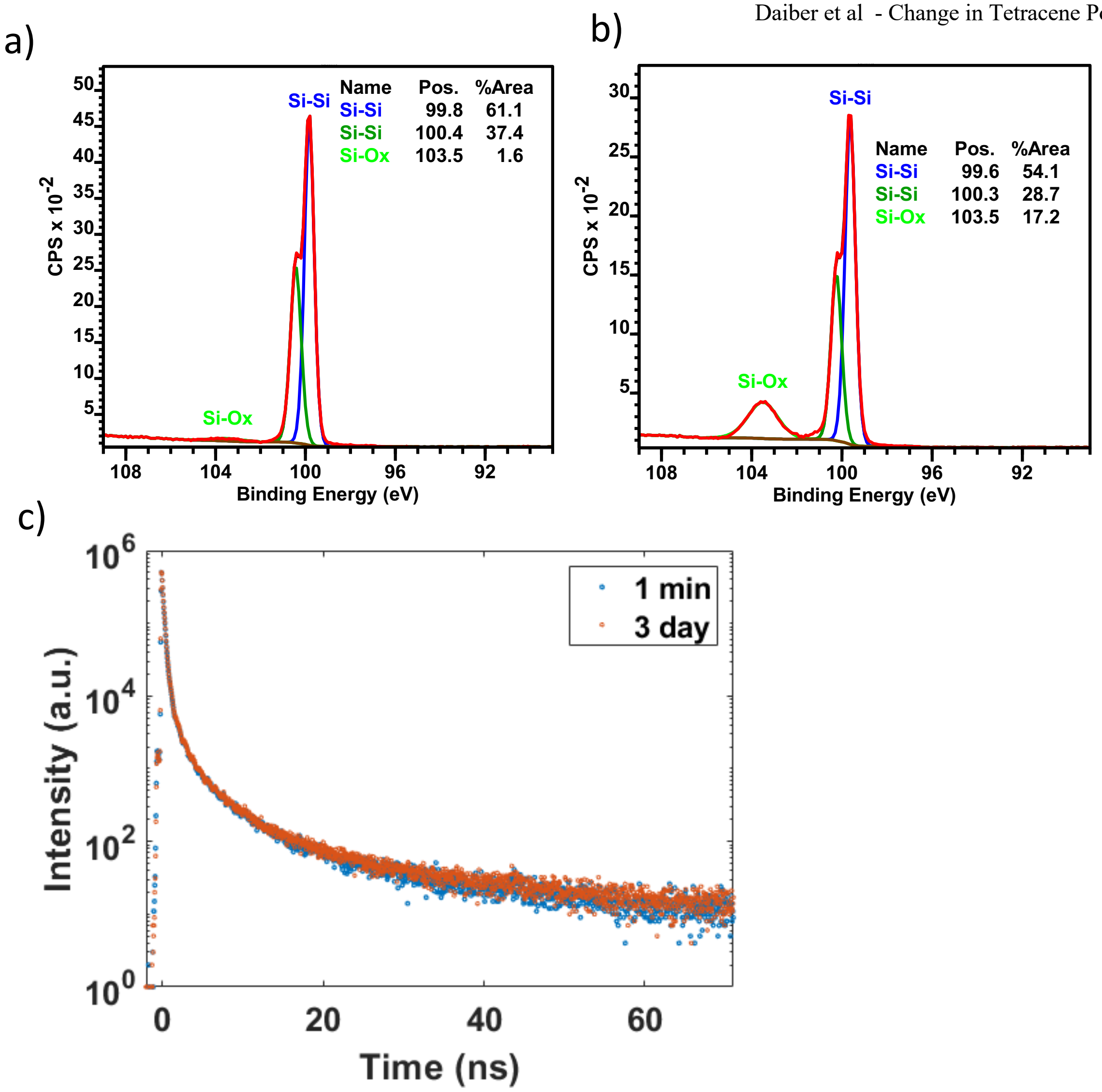

**Figure S7.** Si 2p XPS data of HF-Si sample exposed in air for **a)** 1 min, and **b)** 3 days. The difference shows the growth of $SiO_x$ upon longer exposure. Tetracene was deposited on these pre-exposed HF-Si substrates, and the PL lifetime of tetracene in panel **c)** is unchanged, suggesting exposure of HF-Si to air alone does not lead to triplet transfer.

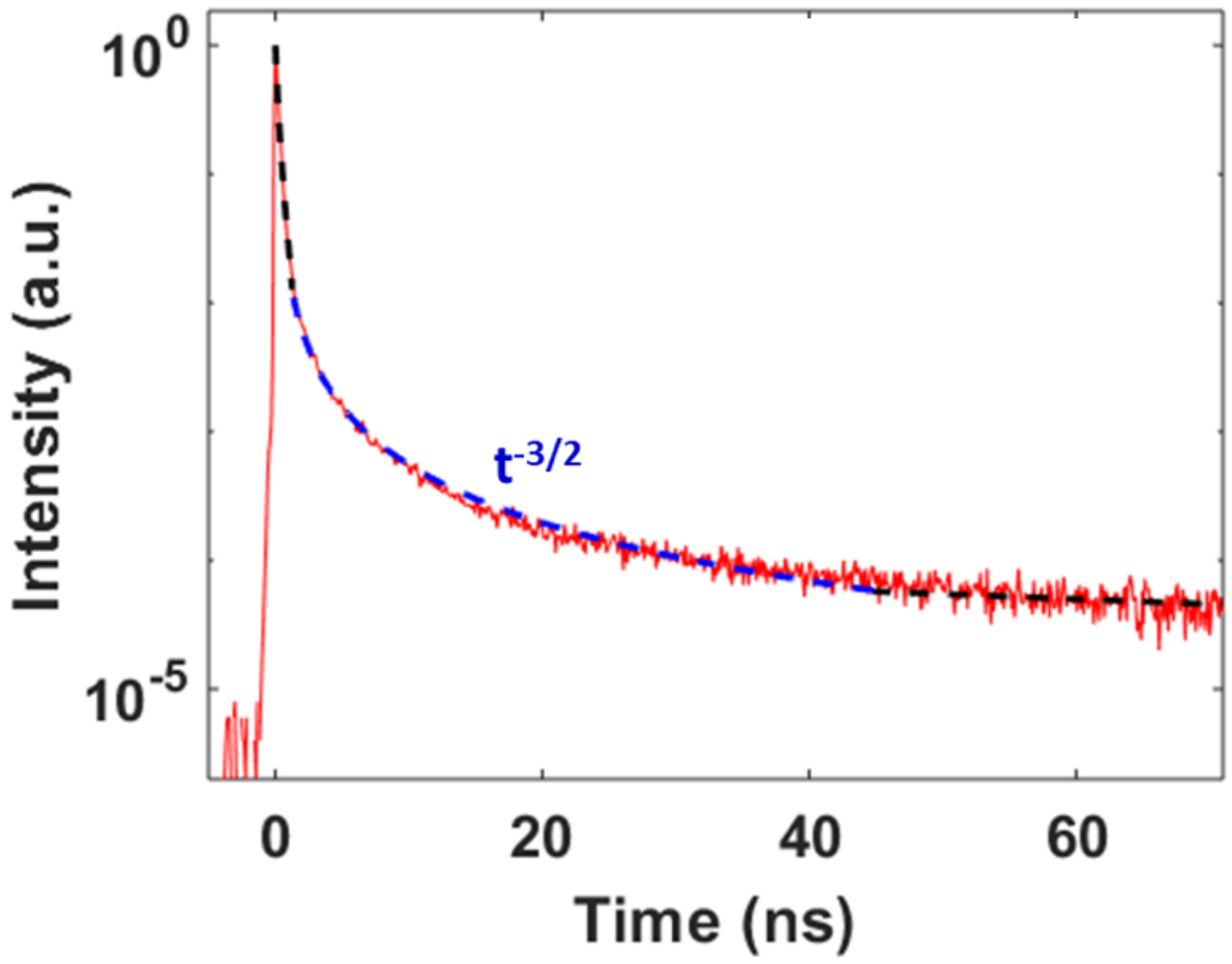

**Figure S8.** The kinetic model based on the solution of coupled differential equations reproduces the data for prompt and delayed PL (black dotted line) of tetracene deposited on $Si/SiO_x$. The intermediate-time region of the PL decay curve is dominated by diffusion of the triplet pair states, which follows a $t^{-3/2}$ dependence.[1] Here the $t^{-3/2}$ term was added empirically to show the match with the data.

**Kinetic Equations for Si/SiO$_2$/tetracene**

$$\frac{d[S]}{dt} = -(K_{SF} + K_{Rad})[S] + K_{TT}[TT]$$

$$\frac{d[TT]}{dt} = K_{SF}[S] - K_{TT}[TT] - K_{Diss}[TT] + K_{TTA}[T]^2 - K_{Trip}[TT]$$

$$\frac{d[T]}{dt} = 2K_{Diss}[TT] - 2K_{TTA}[T]^2 - K_{Trip}[T]$$

$$\frac{d[G]}{dt} = K_{Rad}[S] + K_{Trip}[TT] + K_{Trip}[T]$$

**Kinetic Equations for Si/tetracene**

$$\frac{dS_1}{dt} = -(K_{SF} + K_{Rad})[S] + K_{TT}[TT]$$

$$\frac{d[TT]}{dt} = K_{SF}[S] - K_{TT}[TT] - K_{Diss}[TT] + K_{TTA}[T]^2 - K_{Trip}[TT]$$

$$\frac{d[T]}{dt} = 2K_{Diss}[TT] - 2K_{TTA}[T]^2 - K_{Trip}[T] - K_{TSi}[T]$$

$$\frac{d[T_{Si}]}{dt} = K_{TSi}[T]$$

$$\frac{d[G]}{dt} = K_{Rad}[S] + K_{Trip}[TT] + K_{Trip}[T]$$

The kinetic equations were solved numerically and the solution was plotted against the data.

The singlet ($S_1$) produces a triplet pair (TT) state through singlet fission ($K_{SF}$), which then dissociates ($K_{Diss}$) into free triplets (T). Both singlet, triplet and triplet pair can recombine to the ground state (G). The radiative decay of the $S_1$ state has a lifetime of 12.5 ns (taken from literature, Table S1). The TT state decays back to a $S_1$ state with the rate $K_{TT}$. The free triplets can regenerate the TT state through triplet-triplet annihilation ($K_{TTA}$). The triplet state or triplet pair state decay non-radiatively ($K_{Trip}$) to the ground state. Upon air exposure, triplet transfer to Si also diminishes the triplet population through triplet transfer to Si ($K_{TSi}$).

**Table S1.** Comparison of rate constants determined from the above model with literature for tetracene singlet fission process.

| **Process** | **Our data Si/SiO$_2$/air and HFSi/air** | | **Ref.[2]** | **Ref.[3]** | **Ref.[4]** | **Ref.[5]** |
|---|---|---|---|---|---|---|
| **Expe-riment** | PL decay on polycrystalline tetracene | fs-TA on polycrystalline tetracene | fs-TA microscopy on single crystal | fs-TA on polycrystalline tetracene | PL decay on polycryst. tetracene | |
| **$K_{SF}$** | $(175 \pm 10\ ps)^{-1}$ | same | $(124\ ps)^{-1}$ | $(120\ ps)^{-1}$ | $(90\ ps)^{-1}$ | $(180\ ps)^{-1}$ |
| **$K_{Rad}$** | $(12.5\ ns)^{-1}$ | same | $(524\ ps)^{-1}$ | $(12.5\ ns)^{-1}$ | $(12.5\ ns)^{-1}$ | $(12.5\ ns)^{-1}$ |
| **$K_{TT}$** | $(950 \pm 50\ ps)^{-1}$ | same | $(360\ ps)^{-1}$ | $(1000\ ps)^{-1}$ | $(150\ ps)^{-1}$ | $(100\ ns)^{-1}$ |
| **$K_{Diss}$** | $(280 \pm 15\ ps)^{-1}$ | same | $(439\ ps)^{-1}$ | $(500\ ps)^{-1}$ | $(600\ ps)^{-1}$ | NA |
| **$K_{TTA}$** | $2.2 \times 10^{-11}\ cm^3\ s^{-1}$ | same | 0 | $1.7 \times 10^{-11}\ cm^3\ s^{-1}$ | 0 | 0 |
| **$K_{Trip}$** | $(220 \pm 30\ ns)^{-1}$ | same | $(21.59\ ns)^{-1}$ | $(62.5\ \mu s)^{-1}$ | $(20\ ns)^{-1}$ | $(200\ ns)^{-1}$ |
| **$K_{TSi}$** | - | $(215 \pm 35\ ns)^{-1}$ | - | - | - | - |

Note: TA = transient absorption; PL = photoluminescence.